\documentclass[fleqn,usenatbib]{mnras}

\usepackage{newtxtext,newtxmath}

\usepackage[T1]{fontenc}
\usepackage{ae,aecompl}

\DeclareRobustCommand{\VAN}[3]{#2}
\let\VANthebibliography\thebibliography
\def\thebibliography{\DeclareRobustCommand{\VAN}[3]{##3}\VANthebibliography}

\usepackage{graphicx}	
\usepackage{amsmath}	
\usepackage{bm}

\def\msun{\hbox{M$_\odot$}}
\newcommand{\kms}{km\,${\rm s}^{-1}$}

\usepackage[usenames]{xcolor}
\usepackage[normalem]{ulem}

\title[Exploring NGC1850 BH1]{Updated radial velocities and new constraints on the nature of the unseen source in NGC1850 BH1}

\author[S. Saracino et al.]{S. Saracino$^{1}$\thanks{E-mail: s.saracino@ljmu.ac.uk},
T. Shenar$^{2,3}$,
S. Kamann$^{1}$,
N. Bastian$^{4,5,1}$,
M. Gieles$^{6,7}$,
C. Usher$^{8}$,\newauthor
J. Bodensteiner$^{9}$,
A. Kochoska$^{10}$,
J. A. Orosz$^{11}$,
H. Sana$^{3}$\\
\\
$^{1}$ Astrophysics Research Institute, Liverpool John Moores University, 146 Brownlow Hill, Liverpool L3 5RF, UK\\
$^{2}$Anton Pannekoek Institute for Astronomy, Science Park 904, 1098XH Amsterdam, The Netherlands\\
$^{3}$ Institute of Astronomy, KU Leuven, Celestijnenlaan 200D, 3001 Leuven, Belgium\\
$^{4}$ Donostia International Physics Center (DIPC), Paseo Manuel de Lardizabal, 4, 20018, Donostia-San Sebasti\'an, Guipuzkoa, Spain\\
$^{5}$ IKERBASQUE, Basque Foundation for Science, 48013, Bilbao, Spain \\
$^{6}$ ICREA, Pg. Llu\'{i}s Companys 23, E08010 Barcelona, Spain\\
$^{7}$ Institut de Ci\`{e}ncies del Cosmos (ICCUB), Universitat de Barcelona (IEEC-UB), Mart\'{i} i Franqu\`{e}s 1, E08028 Barcelona, Spain\\
$^{8}$ The Oskar Klein Centre, Department of Astronomy, Stockholm University, AlbaNova, SE-106 91 Stockholm, Sweden\\
$^{9}$ European Organisation for Astronomical Research in the Southern Hemisphere (ESO), Karl-Schwarzschild-Str. 2, 85748 Garching, Germany\\
$^{10}$ Villanova University, Deptartment of Astrophysics and Planetary Sciences, 800 East Lancaster Avenue, Villanova, PA 19085, USA\\
$^{11}$ San Diego State University, Department of Astronomy, 5500 Campanile Drive, San Diego, CA 92182-1221, USA
}
\date{Accepted 08 March 2023. Received 14 February 2023; in original form 16 December 2022}
\pubyear{2022}

\begin{document}
\label{firstpage}
\pagerange{\pageref{firstpage}--\pageref{lastpage}}
\maketitle
\begin{abstract}
A black hole candidate orbiting a luminous star in the Large Magellanic Cloud young cluster NGC 1850 ($\sim100$~Myr) has recently been reported based on radial velocity and light curve modelling. Subsequently, an alternative explanation has been suggested for the system: a bloated post-mass transfer secondary star (M$_{\rm initial} \sim 4-5$~\msun, M$_{\rm current} \sim 1-2$~\msun) with a more massive, yet luminous companion (the primary). Upon reanalysis of the MUSE spectra, we found that the radial velocity variations originally reported were underestimated ($K_{\rm 2,revised} = 176\pm3$~km/s vs $K_{\rm 2,original} = 140\pm3$~km/s) because of the weighting scheme adopted in the full-spectrum fitting analysis. The increased radial velocity semi-amplitude translates into a system mass function larger than previously deduced ($f_{\rm revised}$=2.83~\msun vs $f_{\rm original}$=1.42~\msun). By exploiting the spectral disentangling technique, we place an upper limit of 10\% of a luminous primary source to the observed optical light in NGC1850 BH1, assuming that the primary and secondary are the only components contributing to the system. Furthermore, by analysing archival near-infrared data, we find clues to the presence of an accretion disk in the system. These constraints support a low-mass post-mass transfer star but do not provide a definitive answer whether the unseen component in NGC1850 BH1 is indeed a black hole. These results predict a scenario where, if a primary luminous source of mass M $\ge 4.7$~\msun\, is present in the system (given the inclination and secondary mass constraints), it must be hidden in a optically thick disk to be undetected in the MUSE spectra.
\end{abstract}


\section{Introduction}

Recently, \citet{Saracino2021} reported the discovery of a black hole (BH) candidate orbiting a luminous star in the massive young ($\sim100$~Myr) star cluster NGC~1850, in the Large Magellanic Cloud. Based on the measured radial velocity and luminosity variations of the observed source, and its position in the colour-magnitude diagram (CMD), the authors concluded that the source is a main-sequence turn-off (MSTO) B-type star ($M\sim4.9$~\msun) and that the unseen companion is an $\sim11$~\msun\ BH. Furthermore, the authors suggested that the system is in a semi-detached configuration meaning that the luminous star is beginning to fill its Roche Lobe (they also studied the case of a detached configuration). The system does not display obvious emission lines in the optical region of the spectrum (although the presence of nebular contamination combined with the low spectral resolution of the MUSE observations makes this analysis complicated). However, a faint but significant X-ray detection appears at the position of the source. The lack of a persistent X-ray emission from NGC1850 BH1, although surprising, does not in itself exclude the presence of a BH in the system. Low-mass X-ray binaries with both persistent and transient X-ray emissions are indeed known in the literature (e.g., Cyg X-2, \citealt{CygX2} and V404 Cyg, \citealt{1992Natur.355..614C}, respectively), although neither of them can be directly compared to NGC1850 BH1.

A potential caveat to this discovery is that stars of different masses, which have undergone different evolutionary paths, can display B-type spectra. As an alternative explanation for this system, \citet{El-Badry2021NGC1850} and \citet{stevance2022} have suggested that NGC1850 BH1 is a post-mass transfer binary system, with the brighter source a bloated stripped star with a current mass of $\sim1-2$~\msun\ (M$_{\rm initial}\sim5$~\msun) and the fainter source a more massive star that has gained a lot of mass from the companion (M$_{\rm current}\sim2-5$~\msun). The latter is predicted to be significantly fainter (by approx. 1-2.3 mag in the optical bands) than a main sequence (MS) star of the same mass at the age of NGC 1850 due to rejuvenation episodes occurring during mass transfer \citep{stevance2022}, but see \citet{Wang2020} for an alternative discussion on the impact of mass transfer on the luminosity of the mass gainer. 

We note here that there is a precedence for preferring such a configuration, as previously suggested stellar-mass BH candidates LB-1 \citep{liu2019} and HR~6819 \citep{rivinius20} appear to be best explained instead as post-mass transfer binary stars with two luminous companions \citep[e.g.,][]{Shenar2020LB1, Bodensteiner2020HR, el-badry21}. One important difference, however, between the LB-1 and HR~6819 systems compared to NGC1850 BH1, is that the former systems contain Be stars, i.e., fast rotating B-type stars that display prominent emission lines, while no similar emission is observed in the latter case (see \citealt{kamann2022} for a detailed study of the sample of Be stars in NGC 1850).

Additionally, \cite{El-Badry2021NGC1850} noted an inconsistency in the \citet{Saracino2021} interpretation, namely that if the system is in a semi-detached configuration, then a 5~\msun~ MSTO star would be more luminous than permitted by the observed photometry. In the detached configuration, its implied radius would instead be smaller than the Roche radius, and seems inconsistent with the photometric variability suggesting a (near) Roche filling donor.
On the other hand, in the post-mass transfer model for NGC1850 BH1, we must be catching the system at a unique time, specifically as it is transferring across the Hertzsprung-Russell (HR) diagram from a cool bloated star to a hot sub-dwarf state. The rarity of catching such a system at this time is highlighted in \citet{stevance2022}, where the authors systematically explored a large grid of pre-computed binary models (including mass transfer) and could only find a matching system by significantly expanding the allowed temperature range of the secondary ($\sim10,000$~K compared to the observed $\sim14,500$~K). The chance of catching the luminous component as it crosses the HR~diagram from cool to hot, directly on the MS is then rather small (approx. 1\% of the lifetime of the system), but in principle easier to be detected in this stage than in the later subdwarf stage \citep{Bodensteiner2020HR}.

Upon further modelling of the NGC1850 BH1 system, we uncovered a systematic bias in the published radial velocity measurements. This bias, which will be accurately described in the following Sections, resulted in the underestimation of the radial velocity semi-amplitude $K_2$\footnote{To avoid any confusion in the reader, we specify here that throughout the paper, the observed star is labelled with index 2 and called secondary, while the unseen (more massive) object is labelled with index 1 and called primary.} of the visible source which in turn resulted in an underestimated mass function for the system. In the present work we discuss the updated radial velocity measurements in Section \ref{sec:radial} along with the implications on the estimated orbital properties of the system, especially the mass function. In Section \ref{sec:disentangling} we present upper limits to the presence of a luminous primary stellar component in the system through the technique of spectral disentangling. In Section \ref{sec:min_mass} we focus on the visible secondary star and investigate a plausible lower limit in mass for it. In Section \ref{sec:interpretation} we combine these results and discuss the possible nature of the unseen component based on the new constraints available. Finally, in Section \ref{sec:conclusions} we present our conclusions.

\section{Revised radial velocity and mass function}
\label{sec:radial}

\begin{figure}
\centering
\includegraphics[width=.5\textwidth]{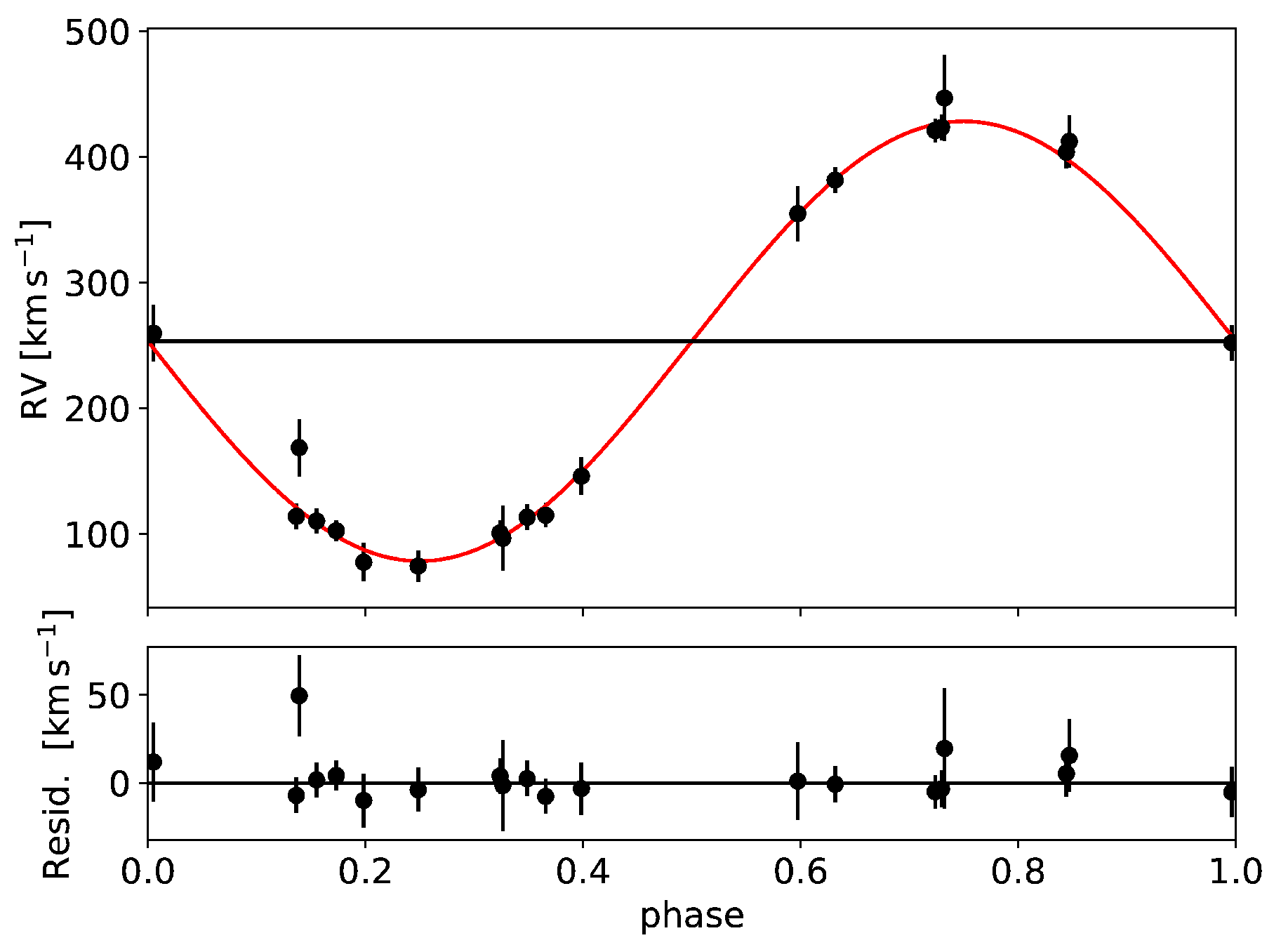}
\caption{The revised MUSE radial velocity curve of the luminous secondary star in NGC1850 BH1, phase-folded using its orbital period of P = 5.04 d, is shown as black dots. The red solid line represents the new best-fitting model (reduced $\chi^2 = 0.52$, RMS = 13.4\,\kms). The bottom panel shows the residuals of the comparison between the observed radial velocities and the best fit model.}
\label{fig:orbitnew}
\end{figure}

Unlike what was done in \citet{Saracino2021}, we present here an alternative method to derive the relative radial velocities of the system, which relies on cross-correlation of the observations with a template spectrum \citep{Zucker1994, Shenar2017, Dsilva2020}. We perform the cross-correlation in the range 7800-8900\,\AA, where several hydrogen lines of the Paschen series are present. As a template, we first use one of the observations themselves. Once a first set of radial velocities has been determined, we compute the co-added spectrum, and use it as a new template for measuring the radial velocities, repeating this process a few times until no notable change in the radial velocities is observed. We convert the relative radial velocities to absolute ones by using the systemic velocity of $V_0 = 253.30\,$\kms~measured by \citet{Saracino2021}. Using this new set of radial velocities we find consistent orbital parameters (e.g. orbital period, eccentricity) to those derived by \citet{Saracino2021}, except for the radial velocity amplitude, which is found to be $K_2 = 175.6\pm2.6$\,\kms. The orbital solution thus derived is shown in Figure\,\ref{fig:orbitnew} while the new single-epoch radial velocities are presented in Table~\ref{tab:muse_data}.

In order to understand the discrepancy in $K_2$ values derived above and reported in \citet{Saracino2021}, who originally found $K_2$ = $140.4\pm3.3$\,\kms, we performed additional full-spectrum fitting analyses using the \textsc{Spexxy} code \citep{Husser2016}, which was used to measure the velocities of the visible star in \citet{Saracino2021}. We found that for this particular star, the weighting scheme used by \textsc{Spexxy} has a significant impact on the measured velocities. By default, \textsc{Spexxy} weighs the spectral pixels by the inverse of their uncertainties during the fitting. If we switch to a more physically motivated inverse-variance weighting scheme, we get radial velocities consistent with those shown in Figure ~\ref{fig:orbitnew}. Using both weighting schemes, we then performed an analysis with \textsc{Spexxy} where we only used the spectral range with $\lambda>7\,800\,\text{\AA}$, in effect using the Paschen series as the only spectral lines in the fit. We found that either weighting scheme (as well as using no weighting at all) resulted again in a velocity curve consistent with the one shown in Figure~\ref{fig:orbitnew}. We repeated the fitting with pPXF code \citep{Cappellari2004,Cappellari2012ascl.soft10002C} for both the entire wavelength range and $\lambda>7\,800\,\text{\AA}$, finding identical results as with \textsc{Spexxy}.

Given the high effective temperature of the observed star, its flux is much higher in the blue part of the MUSE spectral range than in the red \citep[see Figure~2 in][]{Saracino2021}. As a consequence, when using the inverse uncertainties as weights, the blue part (with the strong H$\beta$ and H$\alpha$ lines) has a larger impact on the fit than the red part (containing the Paschen series). Ideally, this over-weighting of the blue part would not affect the kinematics, as all lines are shifted by the same radial velocity. In the case of NGC~1850, however, the strong nebular emission, associated with the 5 Myr old cluster NGC 1850B, represents an additional complication for the data analysis. As both H$\beta$ and H$\alpha$ are particularly strong in the nebular line spectrum, it is conceivable that residual nebular emission contaminates the two line profiles sufficiently so that the velocity estimates based on them are biased towards the cluster mean. An alternative, more physical explanation, could be that accretion onto the unseen companion creates some H$\alpha$ (and potentially H$\beta$) emission that partially fills up the absorption line. However, as the emission component would follow the motion of the unseen primary and hence appear with a phase shift of 180$^{\circ}$ in velocity space, one would expect an over- rather than an under-estimation of the $K_{2}$ value derived using the blue part of the MUSE spectral range (see the discussion in \cite{Abdul-Masih2020} for the LB-1 system).

When comparing the velocity curve shown in Figure~\ref{fig:orbitnew} to the one depicted in Figure~5 of \citet{Saracino2021}, one can see that the scatter of the individual velocity measurements around the Keplerian model is significantly reduced when the revised measurements are used (reduced $\chi^2$ is 0.52). Given this improvement, and the potential issues regarding the usage of the H$\beta$ and H$\alpha$ lines, we give preference to the new result. The difficulty in determining $K_2$ discussed here highlights the need to study the system at higher spectral resolution over a broad wavelength range. This would allow: 1) for a better cleaning of the nebular emission lines (which would be significantly narrower in high-resolution data), also thanks to strong metal lines in the blue that do not appear in the nebular emission; 2) to add stricter constraints from the spectral disentangling technique; and 3) to increase the chances of finding a potential emission-line contribution from accretion onto the unseen companion.

\subsection{Mass Function}

A change in the derived semi-amplitude velocity $K_{2}$ of the visible source in NGC1850 BH1 has a direct effect on the mass function of the system, even if all other orbital parameters (e.g. period, eccentricity) stay the same. In fact, based on the formula of the binary mass function \citep{RemillardMcC2006}, that can be expressed in terms of observational quantities as: 
\begin{equation}
f = \frac{P_{\rm orb}K_{\rm 2}^{\rm 3}(1-e^{2})^{3/2}}{2\pi G},   
\end{equation}
which does not make any assumptions on the mass of the visible source, we obtain $f = 2.83^{+0.14}_{-0.12}$~\msun, significantly higher than $f = 1.42$~\msun~ as derived in \citet{Saracino2021}. Orbital period $P_{\rm orb}$ and eccentricity $e$ are the same as in Table 2 of \citet{Saracino2021}. This implies that, regardless of the mass of the visible star (a normal MS star vs a bloated star), the unseen primary companion is substantially more massive than previously predicted. All the revised and relevant properties of NGC1850 BH1 are listed in Table \ref{table:binary_results}, to provide the reader with a clearer reference.

By using Kepler's third law, the binary mass function can also be written in the form:
\begin{equation}
f = \frac{M_{\rm 1}^{3}\sin(i)^3}{(M_{\rm 1}+M_{\rm 2})^2},   
\label{eq:fm2}
\end{equation}
where $M_{\rm 1}$ and $M_{\rm 2}$ are the masses of the primary unseen component and the secondary visible star, respectively, and $i$ the inclination of the system with respect to the line of sight. This formula suggests that once the mass of the visible star and the inclination of the system are known, the mass of the unseen companion can be determined. Unfortunately, these two additional quantities are uncertain in the case of NGC1850 BH1. In Section \ref{sec:min_mass} we will define an alternative way to put constraints on the mass of the unseen source.

\begin{table}
\centering
\caption{Updated radial velocities using the method outlined in Section~\ref{sec:radial}.}
\begin{tabular}{c c c c}
\hline
Time (MJD) & V$_{R}$ (km/s) & $\sigma$ V$_{R}$ (km/s)\\
\hline
58497.08534751 & 379.44 & 10.12 \\
58498.15614836 & 407.7 & 14.1 \\
58498.17027805 & 404.9 & 25.9 \\
58550.02867354 & 115.18 & 14.15 \\
58550.04180262 & 155.2 & 27.7 \\
58553.01788807 & 418.4 & 10.3 \\
58553.03123202 & 442.5 & 38.0 \\
58556.01231551 & 100.4 & 12.4 \\
58556.02541888 & 100.5 & 27.7 \\
59174.27808058 & 252.7 & 14.0 \\
59174.32203504 & 255.1 & 27.1 \\
59175.16932389 & 103.94 & 12.04 \\
59175.29610617 &  81.6 & 17.7 \\
59176.13916114 & 116.8 & 13.7 \\
59176.30463698 & 149.7 & 14.8 \\
59177.30703879 & 365.2 & 25.1 \\
59190.19805799 & 110.6 & 14.3 \\
59201.25318966 & 120.9 & 13.8 \\
59203.14316930 & 428.29 & 11.16 \\
59251.14817479 & 74.4 & 14.8 \\
\hline
\end{tabular}
\label{tab:muse_data}
\end{table}

\begin{table}
 \caption{Revised properties of NGC1850 BH1}
 \label{table:binary_results}
 \centering
 \begin{tabular}{r l}
 \hline\rule{0pt}{2.3ex}
 Period $P_{\rm orb}$ & $5.0402 \pm 0.0004$ d \\
 Velocity semi-amplitude $K_2$ & $175.6 \pm 2.6$ km/s\\
 Barycentric radial velocity $v_0$ & $253.30^{+2.59}_{-2.44}$ km/s \\
 Mass function $f$ & $2.83^{+0.14}_{-0.12}$ \msun\\
 Eccentricity $e$ & $0.029^{+0.010}_{-0.014}$ \\
 \hline
 \end{tabular}
\end{table}

\section{Spectral disentangling}
\label{sec:disentangling}
Based on the newly derived mass function, which points towards a rather massive unseen companion, we used the MUSE spectra available to set an upper limit on how much this object actually contributes light to the total flux of the system. In fact, if the unseen companion is a massive star as suggested by \citet{El-Badry2021NGC1850} and \citet{stevance2022}, it is rather luminous, so it is expected to contribute significantly to the total flux of the system (but see the discussion about the rejuvenation factor in Section \ref{sec:min_mass}). If the unseen companion is instead a compact object (such as a BH) as suggested by \citet{Saracino2021}, it does not contribute to the light of the system at all if there is no accretion disk around it, regardless of its mass. To do this test, we employed the shift-and-add spectral disentangling technique \citep{Marchenko1998, Gonzalez2006, Shenar2020LB1, Shenar2022SB1s}, which was successfully used to uncover hidden companions in other SB1 binaries (e.g., LB-1, \citealt{Shenar2020LB1}; HR~6819, \citealt{Bodensteiner2020HR}; 28 O-type binaries, \citealt{Shenar2022SB1s}), which have companions contributing as little as $\approx 1-2\%$ to the visual flux.

Briefly, spectral disentangling is the separation of composite spectra into the component spectra of multiple systems, usually performed simultaneously to the derivation of the orbital parameters \citep{Hadrava1995, Bagnuolo1991, Mahy2012}. For given orbital elements, the shift-and-add method relies on an iterative procedure that uses the disentangled spectra obtained in the $j^{\rm th}$ iteration, to calculate the disentangled spectra for the $j+1$ iteration through consecutive shifting-and-adding of the spectra. By minimizing $\chi^2$ between the added disentangled spectra and the observations, one can derive the orbital elements; we refer to \citet{Shenar2020LB1, Shenar2022SB1s} for details. Here, we fix the orbital parameters to those given in Table 2 of \citet{Saracino2021}, except for the radial velocity amplitudes $K_1, K_2$, which are used to minimise $\chi^2$. We note that the light ratio of the components cannot be derived from the disentangling procedure. The adopted light ratio only impacts the final scaling of the spectra.

\begin{figure}
\centering
\includegraphics[width=.5\textwidth]{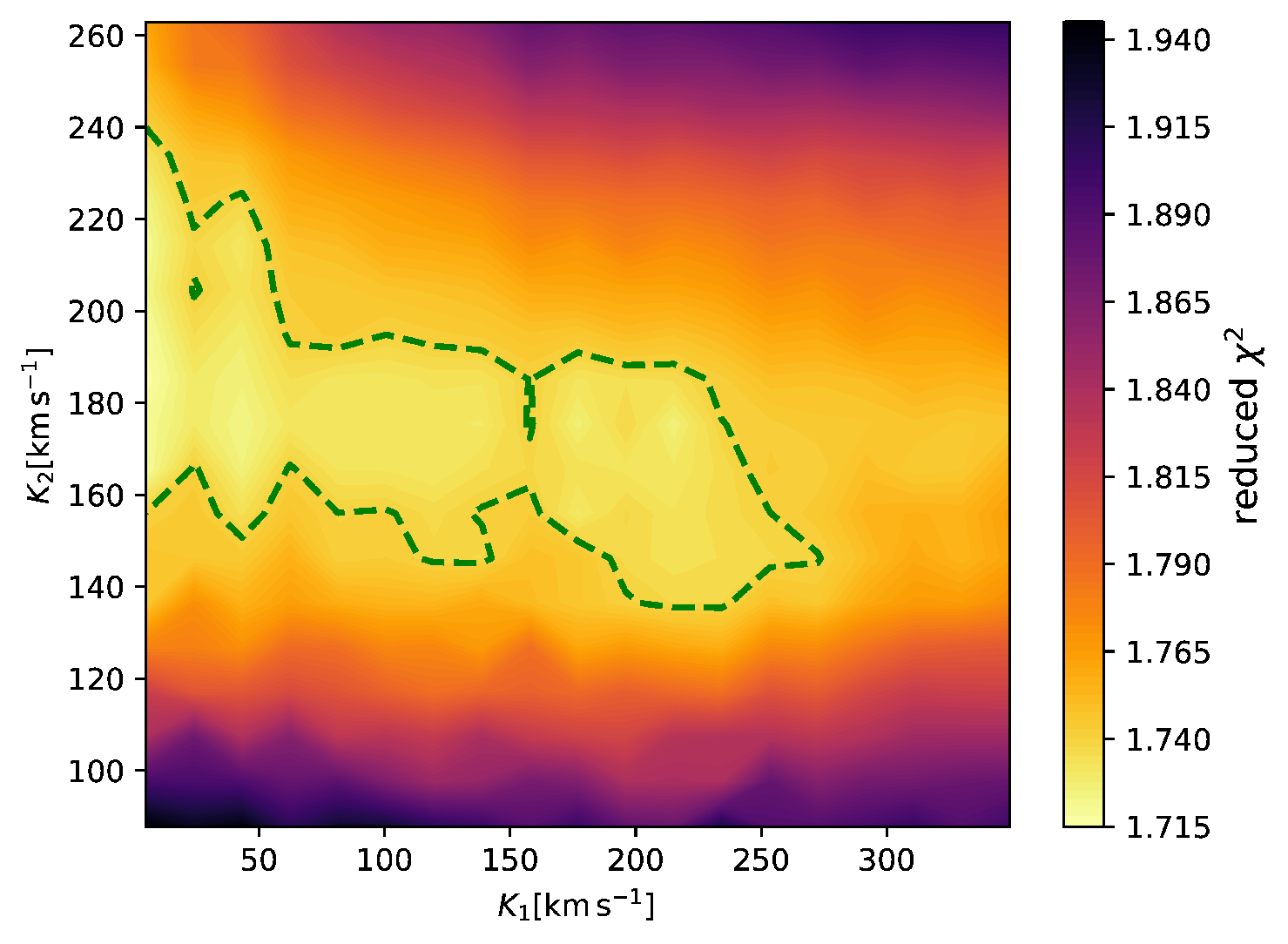}
\caption{$\chi^2$($K_{1}, K_{2}$) from disentangling the spectra in the wavelength region from 8570 to 8910 $\text{\AA}$. A slight correlation between $K_{\rm 1}$ and $K_{\rm 2}$ is observed (see the dashed green line).}
\label{fig:2Dcont}
\end{figure}

In Figure\,\ref{fig:2Dcont}, we show the reduced $\chi^2$($K_{1}, K_{2}$) map obtained when disentangling the four Paschen members (members 8 - 11) in the region $8570-8910\text{\AA}$. Evidently, $K_2$ can be reasonably well constrained and is consistent with the radial velocity measurements in Table \ref{table:binary_results} to within 1$\sigma$. In contrast, the value $K_1$ is poorly constrained, virtually ranging across the entire plausible range of values. We note that disentangling generally yields much larger formal errors compared with standard radial velocity fitting due to the freedom in varying each pixel in the disentangled spectrum. In the figure a slight correlation between $K_{\rm 1}$, $K_{\rm 2}$ is observed. This may possibly indicate that there is some contributing signal from a primary star or disk in NGC 1850 BH1, although this contribution is too small to actually be extracted from the noise using the MUSE data. The presence of a putative accretion disk or a luminous primary (see the discussion in Section \ref{sec:interpretation}) could provide the light signal to explain such a trend in the residual map. Alternatively, this correlation could be a spurious result caused by contaminants (e.g., nebular contamination, tellurics, uncertain normalisation). 

\begin{figure}
\centering
\begin{tabular}{c}
\includegraphics[width=0.47\textwidth]{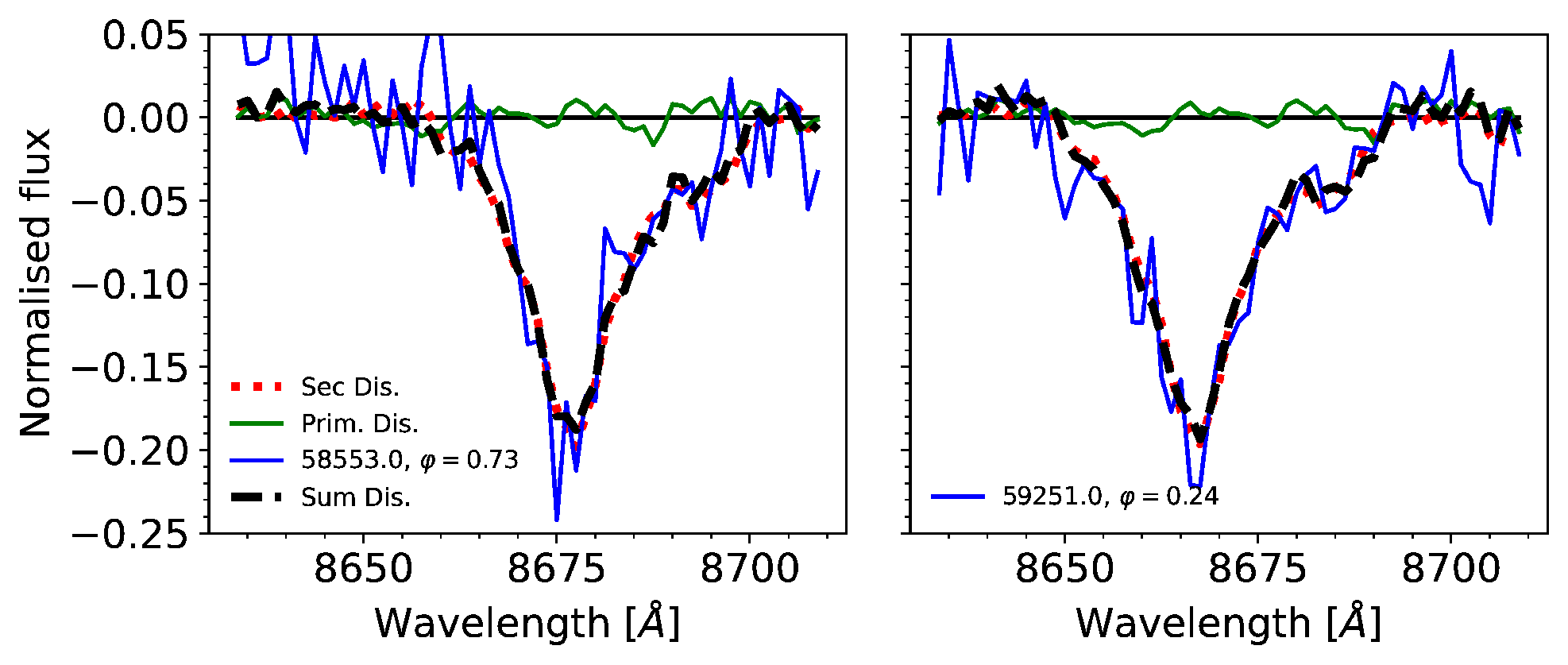} \\
\includegraphics[width=0.47\textwidth]{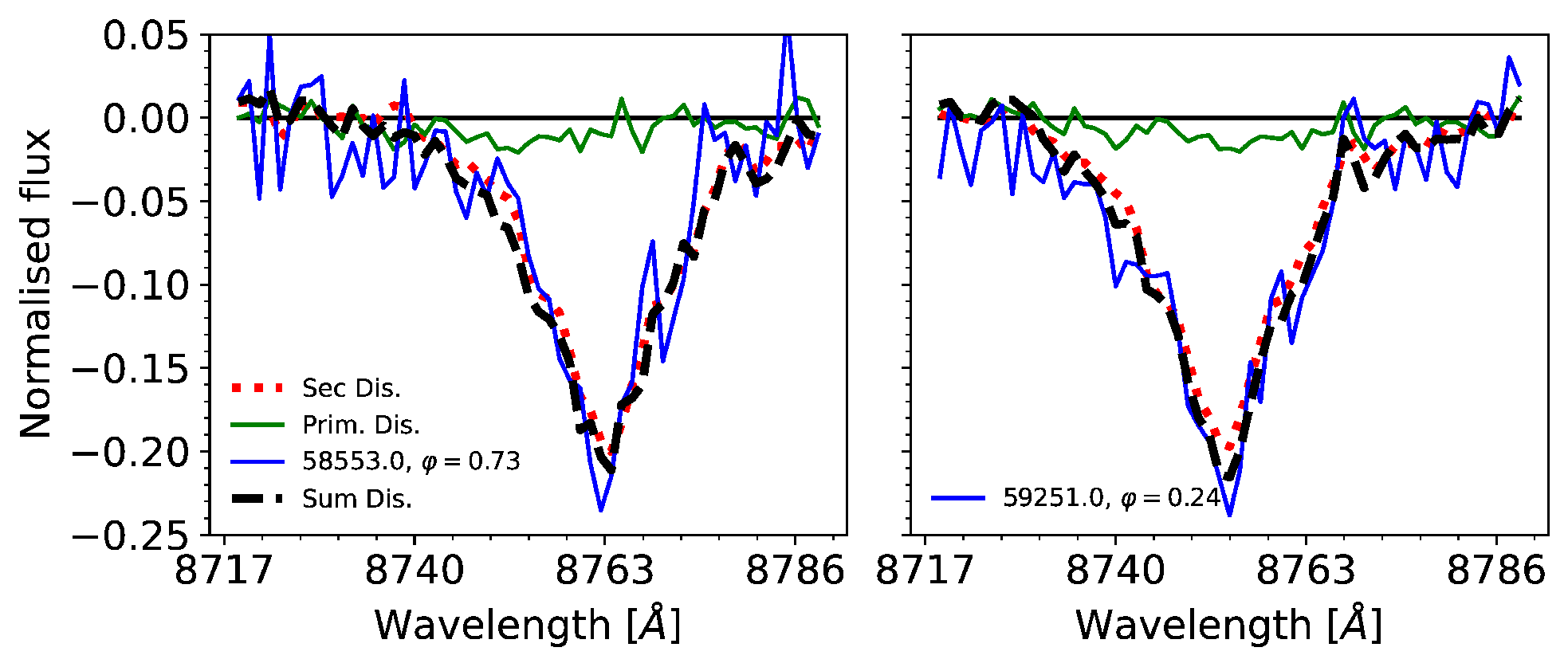} \\
\includegraphics[width=0.47\textwidth]{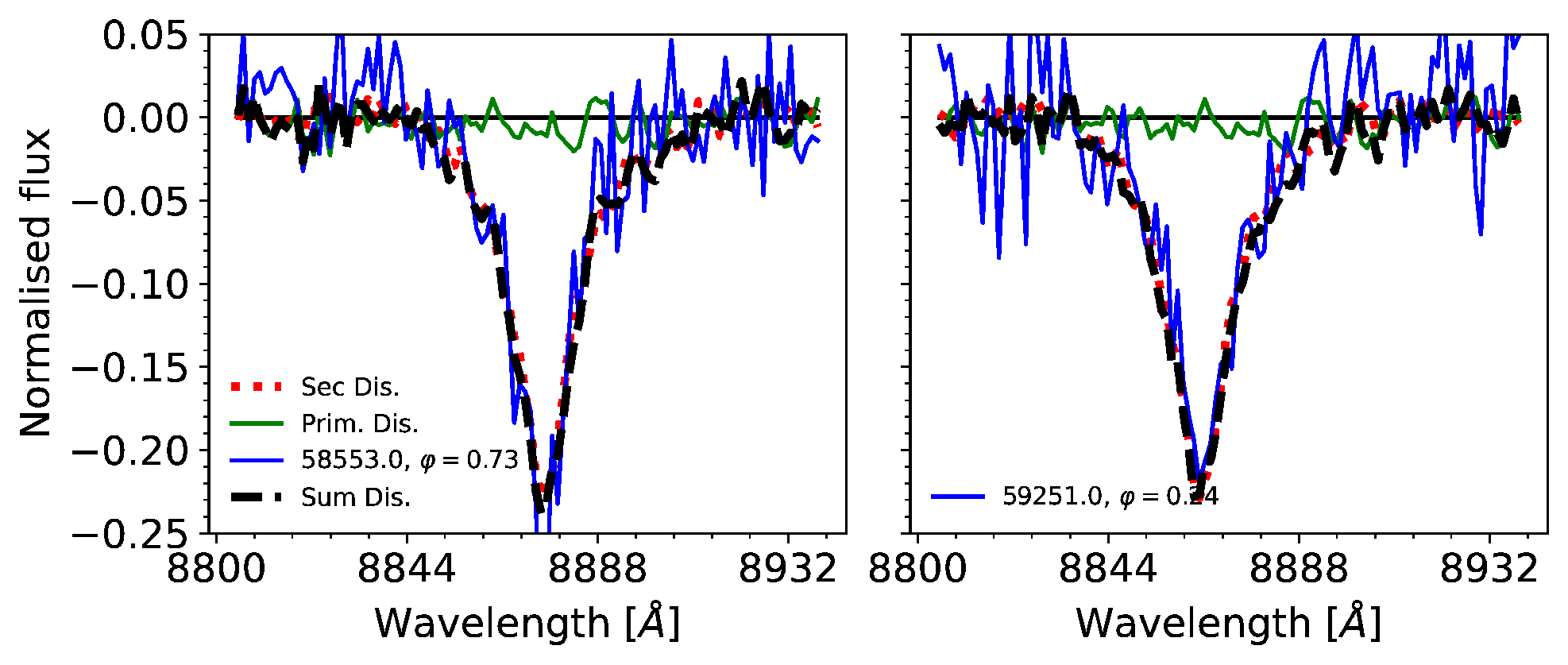} 
\end{tabular}
\caption{
 Disentangled spectra of the primary and secondary for the H\,{\sc i}\,$\lambda 8863, 8750, 8665$ lines (top, middle, and bottom panels) and their sum, compared to observations at radial velocity extremes (left and right panels). The spectra are calculated for $K_2 = 175.6\,$\kms~and $K_1 \approx K_2/4\,$~(41\,\kms). The spectra are not scaled by the light ratio in this figure. The results depend weakly on $K_1$. The spectrum of the primary appears featureless, with the possible exception of the H\,{\sc i}\,$\lambda 8750$.
}
\label{fig:PaschDis}
\end{figure}

Figure\,\ref{fig:PaschDis} shows the disentangled spectra for $K_2 = 175.6\,$\kms~and $K_1 = 41\,$\kms~(i.e., assuming the primary is roughly four times more massive than the luminous secondary). The shifted spectra and their sum are compared to the observations at radial velocity extremes (conjunction phases). Generally, the disentangled spectra of the primary appear close to flat, with the possible exception of H\,{\sc i}\,$\lambda 8750$. We note that the results depend weakly on the adopted value of $K_1$ (values in the range $0.25\,K_2  < K_1 < 4 \,K_2$ were considered). In all cases, the features seen in the spectrum of the primary are comparable to the noise level of the disentangled spectrum.

\begin{figure}
\centering
\includegraphics[width=.48\textwidth]{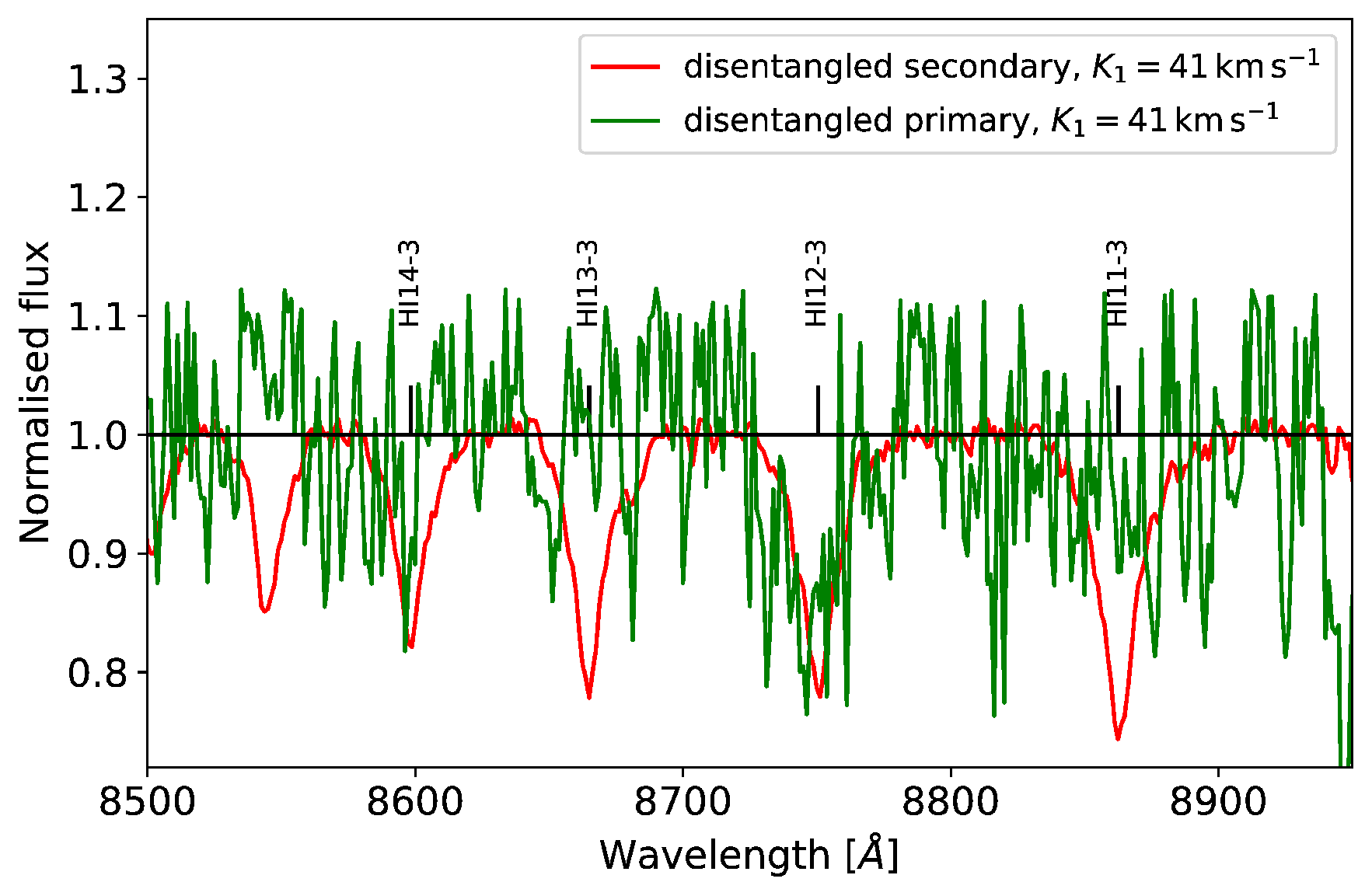}
\caption{Disentangled spectra of the primary and secondary in NGC1850 BH1, obtained for $K_2 = 175.6\,$\kms and $K_1 = 41\,$\kms, and assuming a light contribution of $l_1$ = 10\% for the unseen primary. }
\label{fig:DisSpec}
\end{figure}

In Figure\,\ref{fig:DisSpec}, we show the disentangled spectra of a few neighbouring Paschen lines calculated for $K_2 = 175.6\,$\kms~and $K_1 = 41\,$\kms, assuming a low light contribution for the primary of $l_1 = 0.1$ (i.e., the intrinsic spectrum is multiplied by a factor of 10).
The features that are observed in the disentangle spectra are again of level of the noise, and generally do not overlap with spectral lines. Such features can easily result from non-Gaussian noise, imperfect normalisation, tellurics, or other contaminants. The results imply that, if a non-degenerate companion is present, it must be rather faint. This is corroborated by the simulations below (Section\,\ref{subsec:simu}), where this statement is further quantified.

\subsection{Simulations}
\label{subsec:simu}

\begin{figure}
\centering
\includegraphics[width=.5\textwidth]{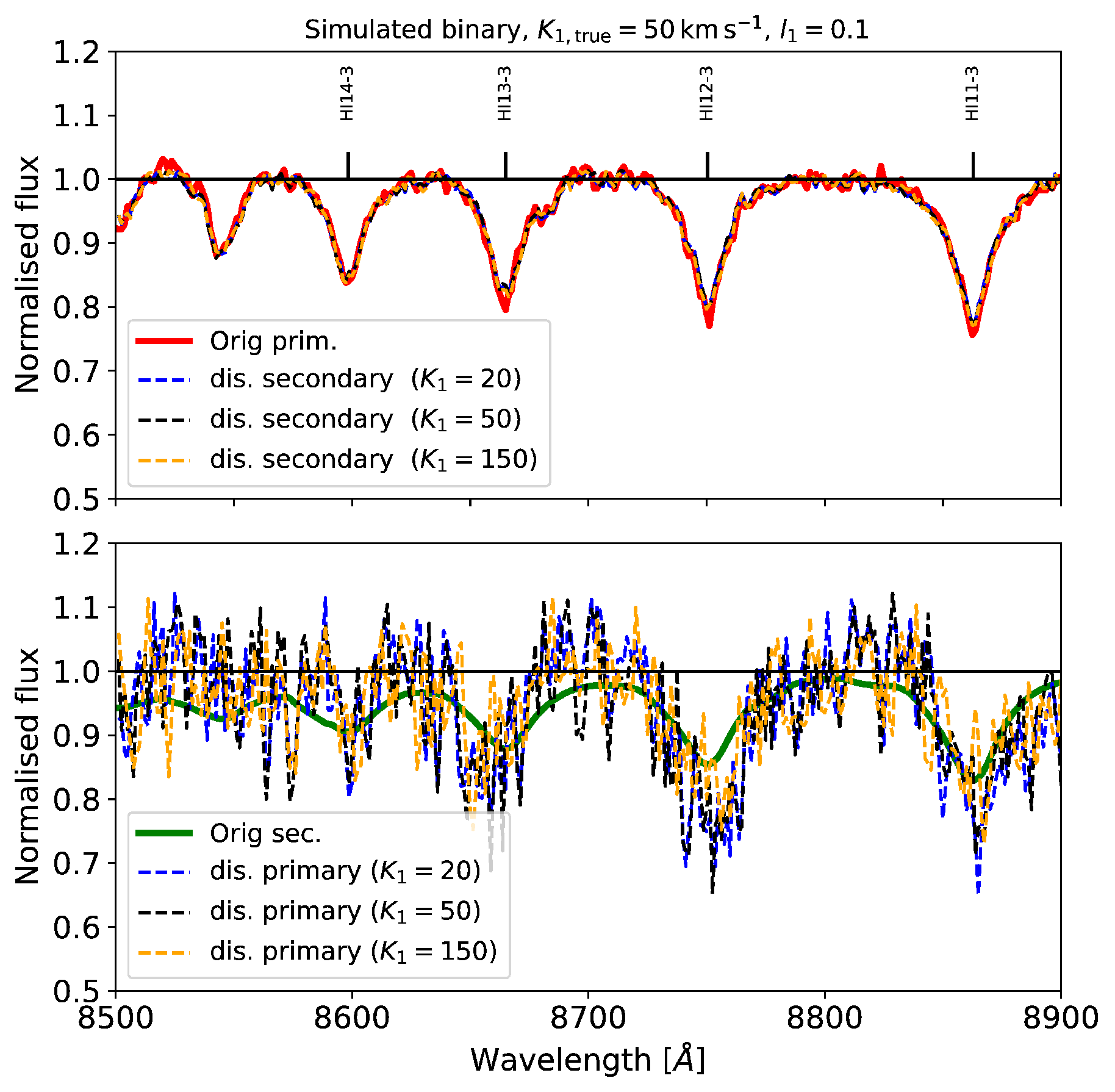}
\caption{The disentangled spectra of the bright secondary (top) and faint primary (bottom) of our simulated binary, compared with the input templates. The disentangling was performed using the same input orbital parameters as used for the simulation, but for $K_1 = 20, 50$, and $150\,$\kms (see legend), illustrating the minor impact of $K_1$ on the spectral appearance of the secondary.}
\label{fig:Sim}
\end{figure}

To test the validity of our method and explore the sensitivity down to which we could detect a hidden MS companion, we simulate a mock binary that mimics the orbit and observational data set of NGC1850 BH1, but contains a non-degenerate companion. For the simulation, we use the co-added spectrum as a template for the luminous secondary. For the mock spectrum of the unseen primary, we use the grids computed with the TLUSTY model atmosphere code \citep{Hubeny1995, Lanz2003, Lanz2007}. We use a model with $T_{\rm eff} = 20,000\,$K, $\log g = 4.0\,$[cgs], and assume that it moves with $K_{\rm 1, true} = 50\,$\kms. To be conservative, we convolve the emergent spectrum of the model with $\varv \sin i = 300\,$\kms~and a macroturbulent velocity of $\varv_{\rm mac} = 30\,$\kms. Finally, motivated by the results in Section\,\ref{sec:disentangling}, we adopt a low light contribution for the primary of $l_1 = 0.1$. The mock observations use the exact S/N values and phases of the original spectra, and are degraded to the MUSE resolution and sampling. 

We then attempted to derive $K_1$ through $\chi^2$ minimisation. However, the $\chi^2$ map is virtually flat, implying that $K_1$ cannot be retrieved. Given the low resolution (compared to the secondary's radial velocities), the intrinsic and rotational broadening of the lines, and the modest S/N, it is not surprising that $K_1$ cannot be retrieved. However, disentangling can still be performed, assuming various values of $K_1$. In truth, the $K_1$ value has a very small impact on the spectral appearance of the disentangled spectra, as long as it is varied in a plausible range. To illustrate this, in Figure\,\ref{fig:Sim}, we show the results from three disentangling experiments of the mock data, varying $K_1$ between 20 and 150\,\kms. The spectra are virtually indistinguishable. This seeming independence on $K_1$ is the result of the broad profiles of the simulated primary and the low spectral resolution of the data.

Evidently, while we cannot retrieve $K_1$, the method yields a spectrum for the hidden primary that matches the original template reasonably well. Some differences are apparent for the primary, which are intrinsic to the method. Since the lines are constantly blended, the disentangling procedure is bound to have some cross-contamination between the stars. However, we note that the differences are boosted by a factor of ten due to the faintness of the primary, such that the deviations seen in Figure \ref{fig:Sim} amount to deviations of the order of a few percent with respect to the mock observations. The exact sensitivity down to which we could detect companions is difficult to establish, since it depends on the stellar parameters, rotation, and light contribution of the primary. However, the experiment described here illustrates that we would very likely be able to detect companions contributing more than $\approx 5-10\%$ to the light.

\begin{figure}
    \centering
	\includegraphics[width=0.47\textwidth]{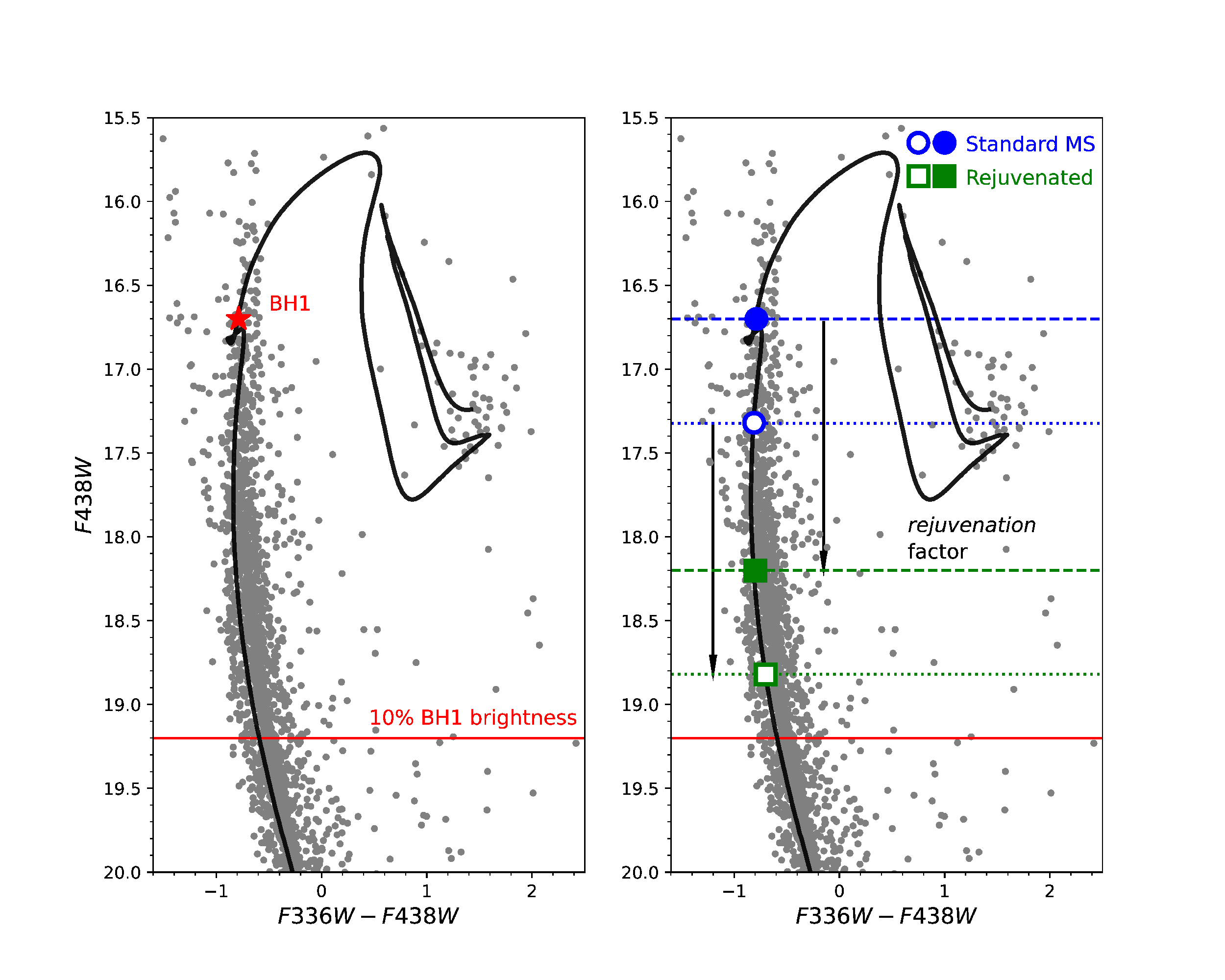}
    \caption{{\it Left panel:} (F336W-F438W, F438W) CMD of NGC 1850, with a MIST isochrone of 100 Myr overplotted. NGC1850 BH1 is represented by a red star in the figure. The red solid line shows the magnitude level corresponding to a brightness of only 10\% of our target (the limit derived by the spectral disentangling). {\it Right panel:} The CMD of NGC 1850 with the same MIST isochrone overplotted. The closed vs open blue dots indicate the magnitude level of a standard (non-interacting) MS star with a mass of 5~\msun (i.e. the mass of a star as bright as NGC1850 BH1) and 4.7~\msun (i.e. the minimum mass for the primary star in NGC1850 BH1, see the text), respectively. The closed vs open green dots instead show how the same two stars look like when the rejuvenation factor due to mass transfer in the binary is applied. They appear $\approx$ 1.5 mag fainter, but still above (by 1 mag and 0.4 mag, respectively) the red solid line which defines the 10\% brightness limit in F438W.}
    \label{fig:lum}
\end{figure}

We visually illustrate how faint the primary (unseen) star must be in terms of magnitude to be undetectable in the MUSE spectra, based on the results of the spectral disentangling. In the left panel of Figure \ref{fig:lum} we present the CMD of NGC 1850 where a MIST isochrone \citep{mist2016} of the appropriate age is overplotted to guide the eye. A red star indicates the position of NGC1850 BH1 (F438W = 16.7, F814W = 16.6) while the red solid line indicates the magnitude level of a MS star (F438W = 19.2, F814W = 19.1) corresponding exactly to 10\% of the brightness of NGC1850 BH1, the limit set by the spectral disentangling.

The conclusions of these tests are twofold: First, if a non-degenerate stellar companion is present in the binary, as suggested by \citet{El-Badry2021NGC1850}, it is fainter than $\approx 10\%$ in the visual. Second, even if a non-degenerate companion is present, it cannot significantly contaminate the spectrum due to its faintness. The stellar parameters determined for the luminous secondary using the co-added observations (which are virtually identical to the disentangled spectrum) should therefore represent the secondary well, unless its light is diluted by an additional sources (e.g., excess emission stemming from a disk). In fact, if there should be an accretion disk in NGC1850 BH1 orbiting around the invisible source, then the results presented above could change to account for this additional component. An extensive discussion of this aspect will be provided in Section \ref{sec:interpretation}.

\section{Minimum mass of the secondary and its implications for the primary}
\label{sec:min_mass}

An additional hint on what is the nature of the invisible source can actually be derived from the analysis of the visible secondary star. The two alternative scenarios that have been proposed to explain NGC1850 BH1 assumed very different masses for the visible source (a normal 5 ${\rm M}_{\odot}$ MS star vs a 1-2 ${\rm M}_{\odot}$ bloated-stripped star). \citet{El-Badry2021NGC1850} argued that the secondary star, if it is indeed filling its Roche lobe, must be less massive than the 5 ${\rm M}_{\odot}$ adopted by \citet{Saracino2021}. According to \citet{eggleton1983}, who derived a formula for the mean density of Roche-lobe filling stars, a 5 $M_{\odot}$ star would be too large (and hence too luminous) to satisfy the photometric constraints. Therefore, we are inclined to adopt the scenario in which the secondary is a low-mass post mass transfer star. Unfortunately, deriving its actual current mass is not possible with the available data, because it is unknown if and how much other sources contribute to the observed magnitudes. Because of this limitation, in this work we define a physically motivated minimum mass for the secondary star and adopt this value in the subsequent analysis. This lower limit directly translates into a minimum mass for the unseen primary star as well, once the inclination of the system is known or can be set to a reasonable value.

From the OGLE light curves available for NGC1850 BH1 and presented in \citet{Saracino2021}, the system does not show any evidence for total or partial eclipses. This might be the case for two reasons: 1) the binary system is made of two luminous sources but it has an inclination such that one source never obscures the other; 2) the unseen source is a dark object (e.g. a BH) so it does not produce eclipses regardless of how the system is inclined. We note here that if the BH is surrounded by an accretion disk, eclipses are still expected unless the system has a geometric configuration similar to that of 1).

According to \citet{beech1989}, in a binary system, the geometric condition for eclipses not to occur is the following: 
\begin{equation}
{\rm \cos}(i) > (R_{\rm 1} + R_{\rm 2}) / a,   
\end{equation}
where $a$ and $i$ are the semi-major axis and the inclination of the system, and $R_{\rm 1}$ and $R_{\rm 2}$ the radii of the primary and secondary stars, respectively. Based on the observational constraints derived from the modelling of the radial velocity curve and the constraints on the radius of the luminous (secondary) star (4.9 ${\rm R}_{\odot} \le R_{\rm 2} \le 6.5 {\rm R}_{\odot}$) imposed by observational uncertainties and the possibility that a second luminous star could contribute to the observed photometry, the lack of eclipses in the OGLE light curves places a limit of $i \leqslant 67^{\circ}$ on the inclination of NGC1850 BH1 (see also \citealt{El-Badry2021NGC1850}). 

\begin{figure}
    \centering
	\includegraphics[width=0.45\textwidth]{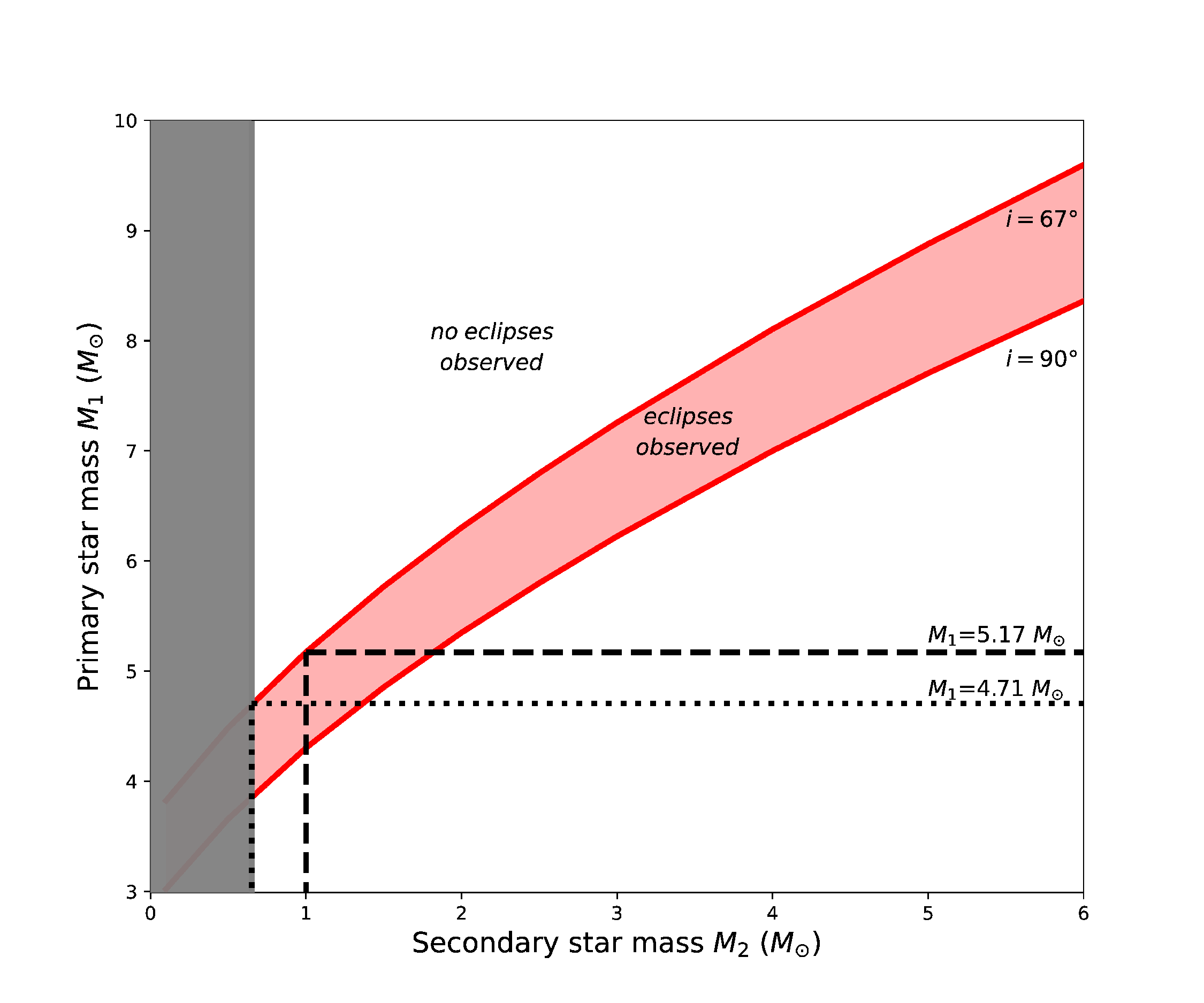}
    \caption{Secondary mass $M_{2}$ vs Primary Mass $M_{1}$ for NGC1850 BH1. The red shaded area defines the region where eclipses are observed, while for $i = 67^{\circ}$ or below (top white area) no eclipses are observed. Our target, NGC1850 BH1, does not show eclipses and assuming $M_{2}=1~$\msun\ as suggested by \citet{El-Badry2021NGC1850}, we obtain a minimum mass $M_{1}=5.17~$\msun\ for the primary (black dashed line). This is also the case for the lowest plausible mass for the secondary star that we set to $M_{2}=0.65~$\msun\ based on what also suggested by \citet{El-Badry2021NGC1850} (dotted line in the plot). The grey shaded area sets the physically motivated lower limit to the mass of the secondary star (see text for details). The white region on the bottom is excluded by the measured binary mass function.}
    \label{fig:constraints}
\end{figure}

In Figure~\ref{fig:constraints} we show, as red solid lines, the mass of the primary (unseen) component as a function of the luminous secondary component, based on the newly measured binary mass function and by adopting equation~(\ref{eq:fm2}) above, for two different inclinations: when the binary is seen edge-on ($i=90^{\circ}$) and when the inclination is $i = 67^{\circ}$, as labelled in the plot. The red shaded area is the region in this parameter space where eclipses are expected to occur, while the white area above the red area is where no eclipses are expected to be observed. In other words, the red solid line at $i = 67^{\circ}$ sets the lower limit to the mass of the primary (as a function of the secondary) if the primary is a star or a BH with an accretion disk. If the primary is instead a dark compact object not surrounded by an accretion disk, the reference line to be considered is the red solid line at $i = 90^{\circ}$.

\citet{El-Badry2021NGC1850} estimate the current mass of the luminous component (a bloated stripped star in their model) to be $\sim1-2$~\msun. In their model, the secondary (initially most massive) component would have recently left the MS and began expanding. During this expansion, the Roche Lobe of the star would have been filled, and mass transfer onto the primary companion would have followed. At the age of NGC 1850 ($\sim100$~Myr), this implies that the initial mass of the secondary would have been $\sim5$~\msun. This is consistent with what \citet{gotberg18} found in their binary interaction models: when assuming an initial mass of $\sim5$~\msun~ for the secondary star, after the mass transfer, the final mass of the stripped star turns out to be of $\sim1-2$~\msun. Although a mass range between 1 and 2~\msun~ is in agreement with current binary evolution models (e.g. \citealt{gotberg18}), it is worth mentioning here that there is not yet enough information about the binary system to allow us to assign a value to the present mass of the visible star. What we do here instead is test how massive the primary would have to be under the assumption of a given secondary mass (and in particular the value of $1$~\msun\ proposed by \citealt{El-Badry2021NGC1850}).

Indeed, by assuming a current mass of the stripped star of $\sim1-2$~\msun, and by applying the no-eclipse condition, the mass of the unseen component is $>5$~\msun. In particular, for a $1$~\msun\ secondary mass, the primary has a minimum mass of $M_{1}=5.17$~\msun, as also highlighted with a black dashed line in Figure~\ref{fig:constraints}. An accretor star of this mass would be as luminous as the visible star itself, so its contribution should be clearly detectable in the MUSE spectra according to the brightness limit set by the spectral disentangling (see Figure \ref{fig:lum}, left panel). 

Based on the BPASS models \citep{BPASS2016}, \citet{stevance2022} pointed out that in a post-mass transfer system, the star that has gained a considerable amount of mass from the companion does not look like a standard (non-interacting) MS star of the same mass and age (in terms of brightness), but instead experiences an episode of rejuvenation, so that in the end it looks much fainter (by up to approx. 2.3 mag in the optical filters). The uncertainties associated with this process are quite large and different binary models tend to predict different scenarios. For example, by using the MESA binary evolution models \citep{Paxton2015}, \citet{Wang2020} recently found that stars who gained a significant amount of mass from their companions in binaries are systematically brighter and more rapidly rotating than they were pre-interaction. This shows that the rejuvenation factor in binary models is still an open question.

While the real factor is somewhat uncertain, we decide to be conservative here and adopt a value of 1.5 mag (fainter) for the rejuvenation factor in the analysis hereafter. Based on this assumption, a primary of $M_{1}=5.17$~\msun~ is significantly fainter than expected from stellar evolution (F438W$\sim$18.2 vs F438W$\sim$16.7) but well detectable in the spectra, as it is still one magnitude brighter than the 10\% brightness limit of F438W = 19.2 imposed by the spectral disentangling\footnote{Since the 10\% flux limit was determined using Paschen lines ($\lambda>7\,800\,\text{\AA}$), we verified that the same behavior is also observed using the F814W filter. In fact, the 10\% limit corresponds to F814W = 19.1, while a primary as bright as the visible star in NGC1850 BH1 would appear to be F814W = 16.6 + 1.5 = 18.1, still one magnitude above the limit.}. This is shown in the right panel of Figure \ref{fig:lum}, where the position of a standard (non-interacting) MS star as bright as NGC1850 BH1 is shown as a closed blue dot, while the same rejuvenated star as a closed green square, overplotted on the CMD of NGC 1850. The impact of the rejuvenation factor on the brightness of the primary (unseen) source will be illustrated more clearly later in the Section, when we define the minimum mass that a primary mass can assume. 

Both \citet{stevance2022} and \citet{El-Badry2021NGC1850} explored evolutionary scenarios that could produce the binary NGC1850 BH1 using BPASS \citep{BPASS2016} and MESA \citep{Paxton2015}, respectively.
They could reproduce the observational properties of the binary system by assuming that the luminous star was a bloated stripped star of $\sim1$~\msun. While it seems very unlikely that the secondary star is significantly less massive than predicted in their models, we want to be conservative here and set the minimum mass for the visible star in NGC1850 BH1 at $M_{2}=0.65\,{\rm M}_{\odot}$, which is the lower limit that \citet{El-Badry2021NGC1850} derived in their work on the basis of the minimum possible radius ($R_{\rm 2} = 4.9 {\rm R}_{\odot}$) this star can assume given its temperature, its CMD position, and the possible presence of a primary luminous source contributing to the photometry. Moreover, since the observed spectrum of the luminous source contains prominent hydrogen lines and easily resembles that of a normal B-type star, it is reasonable to infer that this source has not yet been completely stripped of its entire hydrogen envelope, hence its mass cannot be very low.

By imposing a lower limit for the secondary mass of 0.65~\msun, we consider any mass below this threshold as non-physical and present it as a grey shaded area in Figure \ref{fig:constraints}. The dotted line in the Figure instead shows that a secondary mass of $M_{2}=0.65~{\rm M}_{\odot}$ directly translates into a mass of $M_{1}=4.71~{\rm M}_{\odot}$ for the primary (unseen) component based on the mass function of the system. Assuming it is a normal MS star (F438W $\sim$ 17.3, F814W $\sim$ 17.25), this mass corresponds to a brightness of $\sim 56$\% of that observed for NGC1850 BH1 in the visual (F438W). If we take into account the rejuvenation factor, which makes this star appear fainter by 1.5 mag, we deduce for it a magnitude F438W = 17.3 + 1.5 = 18.8 (F814W = 18.75), which has $\sim$ 14.5\% of the brightness of NGC1850 BH1. It would be barely but still observable in the spectrum, given the 10\% limit found by the disentangling. To illustrate this, the right panel of Figure \ref{fig:lum} shows the CMD of NGC 1850, with the MIST isochrone appropriate for the cluster. The open blue dot in the Figure represents the F438W magnitude of a primary (unseen) star with the minimum allowable mass, 4.71 ${\rm M}_{\odot}$ (assuming it to be a MS star). The open green square instead shows its position once the rejuvenation factor of 1.5 mag is applied, due to mass transfer in the binary. Even for the lowest possible primary mass (hence faintest), this star is expected to be visible, as it is still brighter (by 0.4 and 0.35 mag in F438W and F814W, respectively) than the brightness limit set by the disentangling (here shown as a red solid line). More massive, rejuvenated, primary stars would appear even brighter, thus even easier to detect in the spectra.

From this analysis we can conclude that, if we want to have a system with two luminous stars, for any reasonable secondary (present-day) mass for the visible star, we must invoke the presence of an additional component in the system which by shielding the light of this massive luminous source causes it to contribute very little to the total light.

\section{The unseen source in NGC1850 BH1}
\label{sec:interpretation}

The combination of i) the measured high binary mass function ($f = 2.83^{+0.14}_{-0.12}$~\msun), ii) the lack of observed eclipses in the OGLE optical light curves, and iii) the lower limit of $M_{\rm 2} = 0.65\,{\rm M}_{\odot}$ set to the mass of the visible star results in the primary (unseen) companion of NGC1850 BH1 to have a mass $M_{\rm 1}>4.71$~\msun. A {standard (non-accreting)} MS star of such a mass would be more than half as bright as the observed system, hence its contribution would be easily detectable in the MUSE spectra of the visible companion. However, when the rejuvenation factor of 1.5 mag is taken into account, the brightness of this source drops from 56\% to 14.5\% brighter than NGC1850 BH1 (see Figure \ref{fig:constraints}, right panel). It is important to note that, although much fainter, it does not go below the 10\% brightness constraint we derived from the spectral disentangling. In other words, the contribution of the primary in this configuration is expected to be faint but still detectable in the spectra. 

To reconcile this result with the fact that we do not observe any contribution from the primary in the analysis of the MUSE spectra, there are two viable possibilities to explore: First, the unseen primary star is a non-luminous compact object, and since its minimum mass is higher than that of any possible neutron star (${\rm M}$$\sim$3 ${\rm M}_{\odot}$; \citealt{NS2001}), it is a BH. Second, the unseen primary star is a rather massive luminous source enshrouded in a thick accretion disk which absorbs part of its optical light, making it undetectable. Which of these two possibilities has to be preferred is unclear to date but the scope of this paper is to present the current knowledge about NGC1850 BH1 and suggest possible ways to distinguish one or the other scenario with further observations.

As already noted by \citet{Saracino2021}, the NGC1850 BH1 system belongs to a class of objects called Double Period Variables (DPVs, \citealt{Mennickent2003}), i.e. it shows two periodicities where one is about 33 times longer than the other. There is not much literature on DPVs, and the origin of the longest periodicity is still unknown, however, the general consensus is that these systems are semi-detached (one of the two components fills its Roche Lobe) and made up of two stars, one of which (the gainer) is typically a B-type star surrounded by an accretion disk. In one specific case, HD 170582, it has been suggested that the gainer is bright and massive and should contribute nearly 50\% of the total system light, but since it is encased in an optically thick disk that almost completely obscures it, it only contributes for about 10\%, thus becoming barely detectable \citep{mennickent2015}. For sake of completeness, it is worth mentioning that the accretion disk ($\sim21\,{\rm R}_\odot$) and the semi-major axis ($\sim61\,{\rm R}_\odot$) deduced for the system HD 170582 are much larger than allowed for the configuration of NGC1850 BH1. This is an example but an in-depth comparison of NGC1850 BH1 with the properties of other DPV systems is beyond the scope of this document.

\begin{figure}
    \centering
	\includegraphics[width=0.38\textwidth]{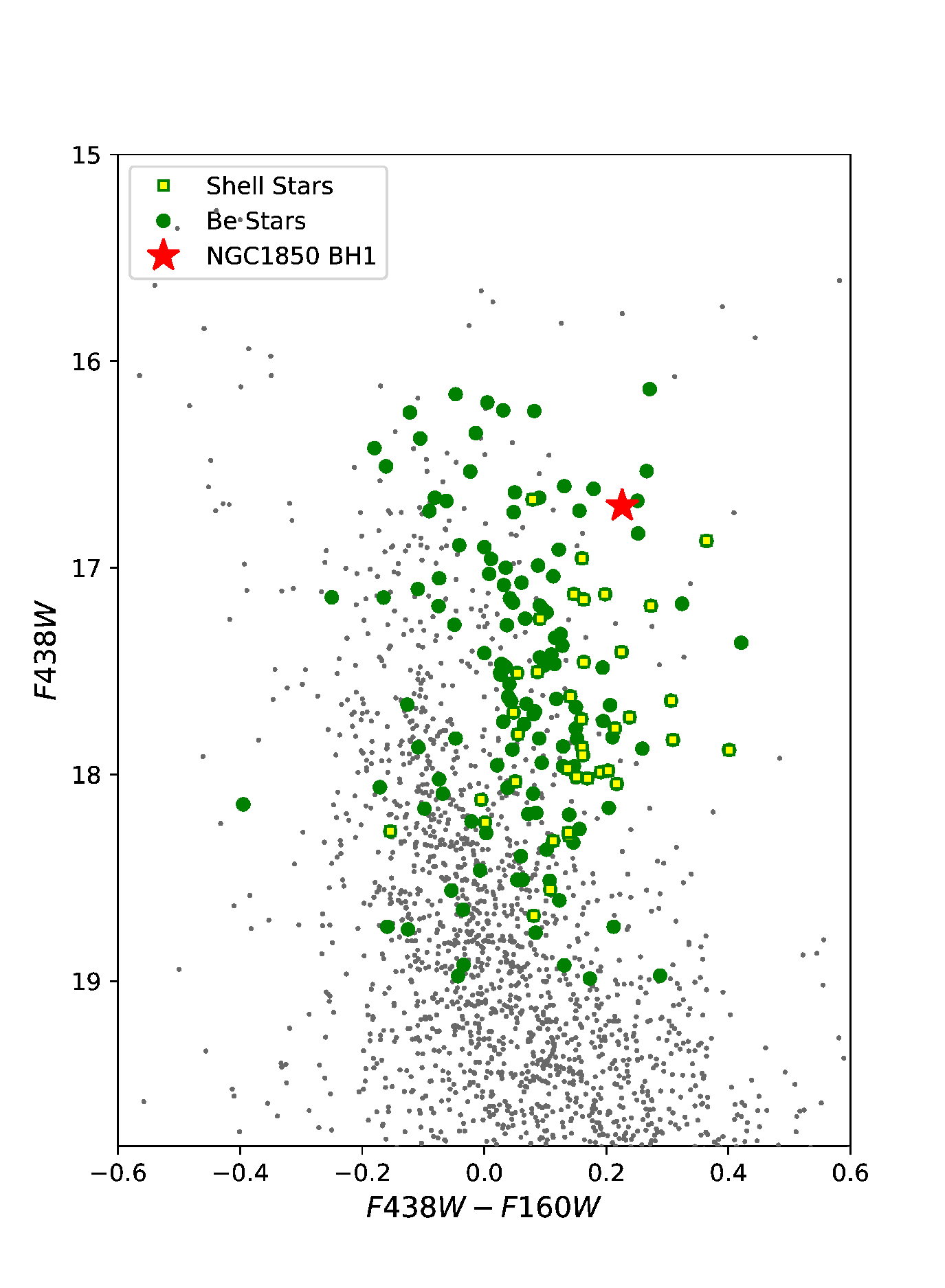}
    \caption{HST/WFC3 near-infrared CMD of NGC 1850. All stars in the cluster are shown as grey dots. Our target, NGC1850 BH1, is instead highlighted as a red star, along with a sample of Be and shell stars spectroscopically identified in the cluster by \citet{kamann2022} and presented as green and yellow dots, respectively. Both Be and shell stars show a near-infrared excess due to disk emission. Although the nature of NGC1850 BH1 is very different, this system exhibits a similar color to many of the shell stars, supporting the idea that an accretion disk is also present in this system.} 
    \label{fig:IRexcess}
\end{figure}

By analyzing archival near-infrared HST/WFC3 data of NGC1850 BH1 (from F105W to F160W), we measured a 2$\sigma$ excess both in F140W and F160W compared to other cluster members in a similar position in the CMD, which seems to support the presence of a third component in the system, namely an accretion disk. Figure \ref{fig:IRexcess} shows an optical/near-infrared (F438W-F160W, F438W) CMD of NGC 1850, where the position of NGC1850 BH1 is presented as a red star. As a comparison, we highlight in green and yellow respectively a sample of Be and shell stars of NGC 1850 studied in \citet{kamann2022}. Shell stars are Be stars (i.e. rapidly rotating B stars) observed (partially) through their disks \citep{Rivinus2006,Rivinius2013}. As shown in the Figure, both Be and shell stars exhibit near-infrared excesses, i.e. they are systematically located on redder colors than normal stars of similar magnitudes. This excess is believed to be mainly caused by emission from their disks. NGC1850 BH1 shares a similar colour, hence a similar near-infrared excess, with many of the shell stars in the cluster so, although the nature of NGC1850 BH1 is very different from that of shell stars, an analogy between these sources can still be drawn. The observed excess of the binary provides further support for the existence of a disk in the system. Additional constraints (e.g. the expected slow rotation of the secondary (luminous) star for a synchronized binary) suggest that, if present, the disk is around the primary (unseen) star.

If the evidence for an accretion disk around the unseen source in NGC1850 BH1 is confirmed by further studies, this will unfortunately still not provide a final answer as to what is the unseen component. Given the high probability that NGC1850 BH1 is a post-mass transfer system, based on recent findings and reasoning, it would be equally plausible to have a disk around a BH or a massive luminous star. The only way to effectively discriminate between the two scenarios is to measure the temperature of the putative disk itself as it is expected to be very different in the two configurations. In particular, if a 5~\msun\ luminous star (and $T\textsubscript{eff}\sim$ 15,000K) is enclosed in an optically thick disk, the light it emits is almost completely absorbed by the disk at optical wavelengths and re-emitted by it at infrared wavelengths. The system thus becomes particularly bright in the near and mid-infrared, given the lower extinction of starlight and the added contribution of the disk, which is expected to be much cooler than the star itself ($T\textsubscript{eff}\ll$ 15,000K). Alternatively, if a BH is part of the system, the properties of the accretion disk are significantly different, with a temperature likely higher than 15,000K near the inner edge, but decreasing with radius as predicted by \citet{ShakuraSunyaev1973}. Near and mid infrared observations of the NGC1850 BH1 system, as those recently secured with the new ERIS/NIX imager at the VLT \citep{Davies2018}, will help to investigate this aspect in more detail. 

\section{Conclusions}
\label{sec:conclusions}

The main results of this study can be summarised as follows:

\begin{itemize}
    \item We have discovered a systematic bias in the measured radial velocities which has been traced to the weighting scheme adopted in the {\sc SPEXXY} results. We have provided updated radial velocity measurements for each epoch. Based on the new modelling of the radial velocity curve, we have updated the radial velocity semi-amplitude to $K_2 = 175.6\pm2.6$\,\kms.
    \item The increased (by 20\%) semi-amplitude velocity thus derived has significantly increased the mass function of the system to $f = 2.83^{+0.14}_{-0.12}$~\msun.
    \item From spectral disentangling we find that only one source is significantly contributing to the spectrum, i.e., any possible stellar secondary contributes at most $10$\% to the optical flux of the system.
    \item The secondary (visible) star is most likely a low-mass post-mass transfer star, but the information available so far does not allow us to assign a value to the present mass of this binary component. Indeed, it is unknown if and how much (other) sources contribute to the observed magnitudes.
    \item Based on the new binary mass function, lack of observed eclipses in the light curves of NGC1850 BH1, and constraints on the luminosity of the system's components, there are two viable possibilities: the unseen component in NGC1850 BH1 is: 1) a BH, its mass being $>3$~\msun\, with the possible addition of an accretion disk; or 2) a bright, rejuvenated star with a minimum mass of $M_{1}\sim4.7$~\msun, enshrouded in an optically thick disk that partially absorbs its light so that it is undetectable in the currently available spectra.
    \item NGC1850 BH1 is a DPV and appears to show an excess in the near-infrared, which can be interpreted as evidence for the presence of a disk in the system. However, both scenarios are still equally likely. Constraining the properties of the disk (e.g. temperature, size) will be one good way to shed more light on the nature of the invisible source.
   \item A scenario in which the primary (unseen) component in NGC1850 BH1 is a BH faces substantial issues regarding its evolutionary history, if we assume a binary origin for it as in the BPASS and MESA models, i.e. the initial period of the binary would be lower than physically allowed given the size of the individual components. A possible caveat of these models, however, is that they only consider isolated binaries and do not include hierarchical triples/quadruples nor the effect of dynamical interactions in clusters. This might instead be appropriate for NGC1850 BH1 which belongs to NGC 1850. In conclusion, since neither the exact configuration of the binary (in terms of $M_{1}$, $M_{2}$, mass ratio etc.) nor its evolutionary history are known, we unfortunately cannot draw any definitive conclusion on this aspect.
\end{itemize}

In a future study we will present detailed modelling of the OGLE light curves of NGC1850 BH1, which also includes the presence of an accretion disk. This will provide stricter constraints on the nature of both the luminous secondary star and the unseen primary companion in the system, sensibly limiting the parameter space we can move in. Moreover, this work clearly shows the urgent need for further and more detailed studies of this peculiar binary system, NGC1850 BH1. They would be helpful to investigate a few important but still unknown aspects:
First, high resolution spectroscopy with a wide wavelength coverage will be essential 1) to apply the disentangling technique in order to be able to detect companions contributing as little as $\approx 1-2\%$ to the visual flux of the system; 2) to study the properties of the luminous (secondary) component (e.g. surface gravity, rotational velocity, chemical abundances); 3) to assess whether the putative disk, if present, dilutes the companion at all bands in a similar way; 4) to place unprecedented constraints on the rejuvenation episodes that occur in binary systems when one of the two sources gains a significant fraction of mass from the companion. The rejuvenation factor is a very uncertain parameter and limiting its allowed range would be a great achievement for future binary evolution studies.
Secondly, near-infrared high resolution photometry will be important to investigate the detailed properties of the putative accretion disk in the system, for example in terms of radius and temperature.

Those mentioned above are essential steps in deciphering the properties of the unseen source in NGC1850 BH1. 

\section*{Acknowledgements}

We thank the anonymous referee for the careful reading and analysis of the paper. We thank Kareem El-Badry and H{\'e}lo{\"i}se F. Stevance for the insightful discussions on the system. We are also thankful to Matti Dorsch and Ulrich Heber for their contributions to the analysis of the MUSE spectra.
SS acknowledges funding from STFC under the grant no. R276234. TS acknowledges support from the European Union's Horizon 2020 under the Marie Skłodowska-Curie grant agreement No 101024605. SK acknowledges funding from UKRI in the form of a Future Leaders Fellowship (grant no. MR/T022868/1).
MG acknowledges support from the Ministry of Science and Innovation (EUR2020-112157, PID2021-125485NB-C22, CEX2019-000918-M funded by MCIN/AEI/10.13039/501100011033) and from AGAUR (SGR-2021-01069). CU acknowledges the support of the Swedish Research Council, Vetenskapsr{\aa}det. This research has received funding from the European Research Council (ERC) under the European Union’s Horizon 2020 research and innovation programme (H.S., grant agreement 772225: MULTIPLES).

\section*{Data Availability}
The data underlying this work are already publicly available to the community. The updated radial velocity measurements are instead listed in Table \ref{tab:muse_data}.

\bibliographystyle{mnras}
\bibliography{star224} 


\bsp	
\label{lastpage}
\end{document}